# Peculiarities of magnetic ordering in the S = 5/2 two-dimensional square-lattice antimonate NaMnSbO$_4$

Tatyana Vasilchikova,[a] Vladimir Nalbandyan,[b] Igor Shukaev,[b] Hyun-Joo Koo,[c] Myung-Hwan Whangbo,[d,e,f] Andrey Lozitskiy,[a] Alexander Bogaychuk,[g,h] Vyacheslav Kuzmin,[g] Murat Tagirov,[g,h] Evgeniya Vavilova,[i] Alexander Vasiliev,[a,j,k] and Elena Zvereva[a,j,*]

[a] *Faculty of Physics, Moscow State University, Moscow 119991, Russia*

[*] zvereva@mig.phys.msu.ru

[b] *Faculty of Chemistry, Southern Federal University, Rostov-on-Don 344090, Russia*

[c] *Department of Chemistry and Research Institute for Basic Sciences, Kyung Hee University, Seoul 02447, Korea*

[d] *Department of Chemistry, North Carolina State University, Raleigh, NC 27695-8204, USA*

[e] *State Key Laboratory of Crystal Materials, Shandong University, Jinan 250100, China*

[f] *State Key Laboratory of Structural Chemistry, Fujian Institute of Research on the Structure of Matter (FJIRSM), Chinese Academy of Sciences (CAS), Fuzhou 350002, China*

[g] *Institute of Physics, Kazan Federal University, Kazan 420008, Russia*

[h] *Institute of Applied Research, Tatarstan Academy of Sciences, Kazan, Russia*

[i] *Zavoisky Physical-Technical Institute, FRC Kazan Scientific Center of RAS, Russia*

[j] *National Research South Ural State University, Chelyabinsk 454080, Russia*

[k] *National University of Science and Technology "MISiS", Moscow 119049, Russia*

An orthorhombic compound, NaMnSbO$_4$, represents a square net of magnetic Mn$^{2+}$ ions residing in vertex-shared oxygen octahedra. Its static and dynamic magnetic properties were studied using magnetic susceptibility, specific heat, magnetization, electron spin resonance (ESR), nuclear magnetic resonance (NMR) and density functional calculations. Thermodynamic data indicate an establishment of the long-range magnetic order with $T_N \sim 44$ K, which is preceded by a short-range one at about 55 K. In addition, a non-trivial wasp-waisted hysteresis loop of the magnetization was observed, indicating that the ground state is most probably canted antiferromagnetic. Temperature dependence of the magnetic susceptibility is described reasonably well in the framework of 2D square lattice model with the main exchange parameter $J = -5.3$ K, which is in good agreement with density functional analysis, NMR and ESR data.

## I. INTRODUCTION

Currently, low-dimensional (LD) magnetic oxides of alkali and transition metals are actively studied, due to the prospects of their use as electrode materials for the manufacture of lithium-ion and sodium-ion batteries [1-3]. Among them layered (quasi-two-dimensional (2D)) compounds represent the most numerous and interesting class of the materials both from the point of view of practical aspects and remarkable physical properties. Within 2D layers, the magnetic ions may adopt the different networks including triangular, square, honeycomb, kagome etc. [4-8] that provides an excellent platform to study exotic magnetic phenomena [9, 10]. The investigation of these compounds allows one to determine the spin lattices relevant for the discussion of their magnetic properties based on first principles electronic structure calculations.

To a first approximation, the magnetic properties of a 2D square spin lattice, for which the formation of a spin-liquid ground state is predicted, can be described by the nearest-neighbour (NN) spin exchanges, namely, $J_1$ along the side of the square and next-nearest-neighbour (NNN) exchange $J_2$ along the diagonal of the square i.e., the $J_1$-$J_2$ model [8, 11-20]. Depending on the ratio $\alpha = |J_2/J_1|$ and the sign of the $J_2/J_1$ ratio, three different ordered states can be realized, namely, ferromagnetic (FM), Néel antiferromagnetic (NAF) and columnar antiferromagnetic (CAF) states. In contrast to a NAF state, a CAF state is characterized by ferromagnetic alignment along the x and y axis and is realized when the NNN exchange dominates over the NN exchange ($\alpha \gg 1/2$) [21,22]. In addition, around $\alpha \approx 1/2$, either spin-liquid and plaquette valence bond states ($J_1 > 0$) [16, 20] or spin nematic ($J_1 < 0$) [18] ground states are predicted.

A CAF ground state has been realized for the layered vanadium oxides $Li_2VOXO_4$ ($X$ = Si, Ge), [16, 23-25] which are regarded as the first examples of frustrated 2D square quantum Heisenberg antiferromagnets (QHAF). The later studies of the solid solutions $Li_2V_{1-x}OTi_xSiO_4$ (for $0 \leq x \leq 0.2$) by means of $^7$Li and $^{29}$Si NMR, muon spin relaxation (μSR), as well as magnetization [26] revealed that the ratio $|J_2/J_1|$ decreases with $x$, and the magnetic ordering temperature decreases with reducing the spin stiffness [24] by a factor of approximately $(1-x)^2$. In contrast to the case of $Li_2VOXO_4$ ($X$ = Si, Ge), the $J_1$ and $J_2$ exchanges compete in the $S = 1/2$ ($V^{4+}$) QHAF system, $VOMoO_4$, [27, 28] which orders at a markedly higher temperature (~ 42 K) and demonstrates a structural distortion. It is worthwhile to note that theoretical calculations [16, 28] predict a quite strong interlayer exchange for both $Li_2VOXO_4$ ($X$ = Si, Ge) and $VOMoO_4$.

Another extensively studied family of oxides with 2D square lattice are the vanadium phosphates $AA'VO(PO_4)_2$ ($A,A'$ = Pb, Zn, Sr, Ba, Cd). The parent compound $Pb_2VO(PO_4)_2$ was the first example of frustrated square lattice with a NN FM exchange interaction and a stronger diagonal AFM exchange [29-31]. The spin dynamics probed by ESR established that $Pb_2VO(PO_4)_2$ is the first frustrated square-lattice system with spin topological fluctuations of the Kosterlitz-Thouless type [31].

An earlier work on its isostructural analogue, $Zn_2VO(PO_4)_2$, proposed it to be a one-dimensional (1D) spin chain system [32], but later investigations [33, 34] unambiguously excluded the 1D model and revealed the same thermodynamic behavior as found for $Pb_2VO(PO_4)_2$. All members of $AA'VO(PO_4)_2$ family ($AA'$ = Pb, Zn, Sr, Ba, Cd) investigated up to now are located in the CAF region with frustration parameters $-0.9 \leq \alpha \leq -1.9$ [25, 30, 35].

Other $S = 1/2$ QHAFs with 2D square lattice include $Na_{1.5}VOPO_4F_{0.5}$ ($\alpha = -1.8$) [36] and $AMoOPO_4Cl$ ($A$ = K, Rb) ($\alpha > -10$) [37], which are regarded as good candidates for the spin-1/2 $J_1$-$J_2$ square-lattice magnet with columnar AFM order. $Sr_2VO_4$ presents unusual magnetic properties at low temperature [38-41]. To explain these properties, several proposals have been put forward, which include an orbitally ordered phase due to Jahn-Teller distortion [38], an octupolar order driven by spin-orbit coupling (SOC) [42], and a Néel order with mute magnetic moment, where the spin and orbital moments cancel each other [39]. µSR studies have shown inhomogeneous magnetic state [40] with the sizable competition of ferromagnetic and antiferromagnetic correlations preventing the system from developing a long-range ordered phase. Such a spin-liquid/spin-glass state indeed can be suggested, but recent theoretical calculations predict the single-stripe magnetic ordering [41].

At the moment, only a few 2D square lattice compounds with classical spin were mentioned in the literature. In particular, $CuMnO_2$ is such a rare example, which is topologically representative of a frustrated square lattice with $S = 2$. A neutron powder diffraction study has shown that the long-range 3D magnetic ordering occurs simultaneously with a lowering of the monoclinic symmetry to a triclinic one [43].

The present paper is devoted to the investigation of member of 2D square lattice compounds with classical spin, $NaMnSbO_4$. We performed detailed studies of its crystal structure, static and dynamic magnetic properties supported by density functional calculation.

## II. EXPERIMENTAL

### A. Sample preparation and X-ray diffraction (XRD)

The new compound was prepared by conventional solid-state reactions in inert atmosphere to maintain manganese in its low oxidation state of 2+. The XRD measurements were performed using an ARL X'tra diffractometer with Cu $K_\alpha$ radiation, eliminating almost completely all undesirable wavelengths using an Si(Li) solid-state detector. The powder pattern for the structural study was taken with coffee powder admixed to the sample to reduce grain orientation effect. The crystal structure refinement was performed with the GSAS+EXPGUI suite [44, 45]. Details of the preparation, EDX analysis and crystal structure are presented in the Supplemental Material (pdf and cif) [46,47].

**B. Thermodynamic measurements**

The temperature dependences of the magnetization at 0.1 T and 9 T were measured in the temperature range 2 – 300 K by means of a Quantum Design PPMS 9 system. $M(T)$ curves were also recorded at a number of external magnetic fields (from 0.5 to 9 T) in the temperature range 2 – 100 K. In addition, the isothermal magnetization curves were obtained in external fields up to 9 T at various constant temperatures ($T$ = 2, 55, 155 K) after cooling the sample in zero magnetic field.

Specific heat measurements were carried out by a relaxation method using a Quantum Design PPMS system. Data were collected at zero magnetic field and under applied $B$ = 9 T in the temperature range 2 – 250 K.

**C. ESR and NMR**

Electron spin resonance studies were carried out using X-band ESR spectrometer CMS 8400 (ADANI) ($f \approx 9.4$ GHz, $B \leq 0.7$ T) equipped with a low temperature cryostat, operating in the range $T$ = 5 – 300 K. The effective g-factor has been calculated with respect to BDPA (a, g-bisdiphenyline-b-phenylallyl) reference sample with $g_{et}$ = 2.00359.

Nuclear magnetic resonance experiments were carried out using home-build broad-band pulsed NMR spectrometer [48] with Bruker superconducting 0 – 9 T magnet and the cryostat optCRYO105 from RTI Ltd. Measurements were performed in the external magnetic field 3.67 T in the temperature range 35 – 300 K. The spectra were obtained by solid echo pulse sequence ($\pi/2$-$\tau$-$\pi/2$, with $\pi/2$ = 6 μs, $\tau$ = 30 μs) in step-by-step variation of radiofrequency and integration of FFT lines in each point in frequency range ±25 kHz. Spin-lattice relaxation rates were measured at the frequency of maximum spectral intensity by saturation recovery method.

**III. RESULTS AND DISCUSSION**

**A. Crystal structure**

The XRD powder pattern of NaMnSbO$_4$ (Fig. 1) was indexed to an orthorhombic system with the ITO program [49] and was included into the Powder Diffraction File [50]. The systematic absences unambiguously indicated the space group *Pbcn*. This, together with the similarity of the unit cell sizes in the *ac* plane, suggested that NaMnSbO$_4$ has a structure similar to that of LiSbO$_3$ [51] based on a two-layer hexagonal close packing of oxygen atoms. The third lattice parameter, $b$, was ~ 4/3 that of LiSbO$_3$. Therefore, it was quite natural to add the fourth stack of face-sharing octahedra to accommodate the additional component, MnO. This model was verified and refined to reasonably low discrepancy factors. The main results are listed in Tables 1-3 and the resulting stricture was illustrated in Figs. 1 and 2. The bond lengths and bond valence sums are in reasonable agreement with the expected values (Table 3).

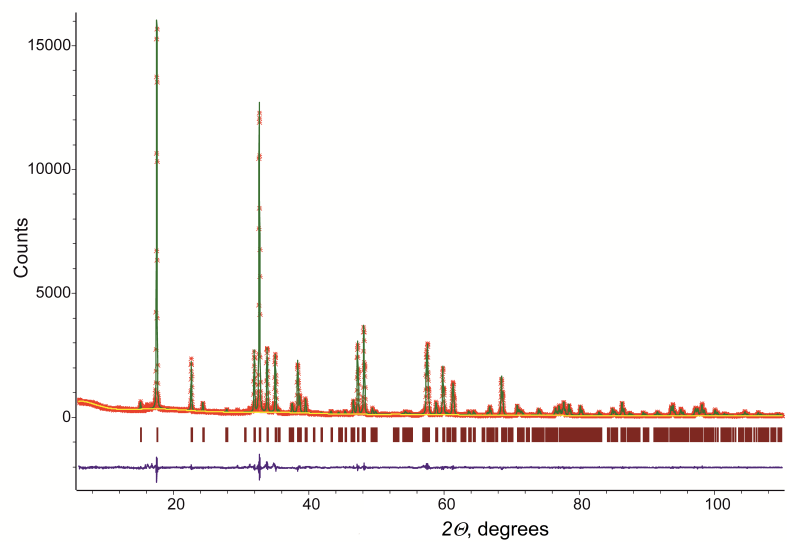

**Fig. 1.** Powder XRD pattern of NaMnSbO$_4$. Asterisks, experimental data; green line, calculated profile; violet line in the bottom, difference profile; vertical bars, calculated positions of Bragg reflections.

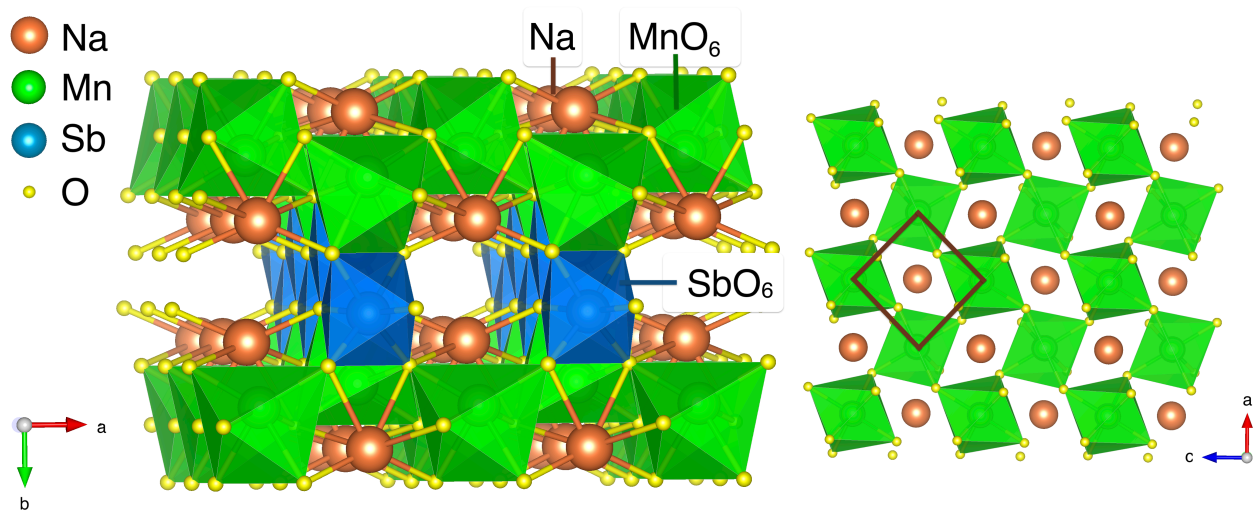

**Fig. 2.** Left part: Polyhedral presentation of the crystal structure of NaMnSbO$_4$. Orange, green, blue and yellow spheres are Na, Mn, Sb and O ions, respectively. The octahedra around the sodium ions are omitted for clarity. Right part: magnetoactive layers of manganese and sodium ions.

From the point of view of magnetic sublattice network in NaMnSbO$_4$, magnetoactive mixed layers of manganese and sodium alternate with nonmagnetic layers of antimony stacked along *b* axis, as shown in Fig. 2. The magnetic Mn$^{2+}$ cations are arranged in an almost square net of vertex-shared octahedra as shown on the right part in Fig. 2 with Mn-Mn distances of 3.8938(5) Å and Mn-O-Mn angles of 128.0(4) Å. The shortest Mn-Mn distances between the adjacent nets are 52 % longer, 5.91Å; thus, the magnetic sublattice seems to be essentially bidimensional. Structural arrangement provides conditions for reducing the dimension of magnetic exchange interactions and frustration of magnetic subsystem.

**Table 1.** Details of the XRD measurements and structure refinement

| | | | | |
|---|---|---|---|---|
| Crystal system | Orthorhombic | Texture parameters | axis 001 | |
| Space group | $Pbcn$(no. 60) | (March-Dollase) | ratio 0.9813 | |
| Lattice constants, Å  a | 5.60043(5) | $2\Theta$ min, ° | 6.02 | |
| b | 11.68849(12) | $2\Theta$ max, ° | 110.00 | |
| c | 5.29327(5) | Step width, ° | 0.02 | |
| Cell volume, Å$^3$ | 346.500(6) | No. of data points | 5200 | |
| Formula weight | 263.67 | No. of reflections calc. ($\alpha_1$ only) | 219 | |
| Z | 4 | No. of variables | 57 (20 – of structure) | |
| Wavelengths, Å  $\alpha_1$ | 1.54056 | Discrepancy | $R(F^2)$ | 0.0411 |
| $\alpha_2$ | 1.54439 | factors | $R_p$ | 0.0655 |
| Ratio | 0.5 | | $R_{wp}$ | 0.0898 |
| Density (calc.), g/cm$^3$ | 5.055 | | $\chi^2$ | 2.360 |

**Table 2.** Atomic coordinates and displacement parameters in NaMnSbO$_4$

| Atom | Site | Symmetry | $x/a$ | $y/b$ | $z/c$ | U |
|---|---|---|---|---|---|---|
| Na | 4c | $C_2$ | 1/2 | 0.1327(5) | 3/4 | 0.0206(17) |
| Mn | 4c | $C_2$ | 1/2 | 0.72602(22) | 3/4 | 0.0083(6) |
| Sb | 4c | $C_2$ | 1/2 | 0.42856(11) | 3/4 | 0.01007[*] |
| O1 | 8d | $C_1$ | 0.3182(10) | 0.5606(9) | 0.5710(12) | 0.0143(21) |
| O2 | 8d | $C_1$ | 0.2990(13) | 0.3150(8) | 0.5987(11) | 0.0140(23) |

[*]Refined anisotropically, $U_{iso}$ is reported

**Table 3.** Important interatomic distances (Å) and bond valence [52] sums (BVS) in NaMnSbO$_4$

| Bonds | Na-O | Mn-O | Sb-O |
|---|---|---|---|
| Distances | 2.187(8)×2 | 2.128(8)×2 | 1.916(9)×2 |
| | 2.540(10)×2 | 2.214(7)×2 | 1.985(6)×2 |
| | 2.566(7)×2 | 2.382(9)×2 | 2.078(9)×2 |
| Average | 2.43 | 2.24 | 1.99 |
| Sums of ionic radii [53] | 2.40 | 2.21 | 1.98 |

| | Na | Mn | Sb | O1 | O2 |
|---|---|---|---|---|---|
| BVS | 1.14 | 1.86 | 4.90 | 2.02 | 1.93 |

## B. Static magnetic properties

The temperature dependence of the magnetization $M$ of NaMnSbO$_4$ demonstrates a complex behaviour with a temperature variation (Fig. 3). At high temperatures it varies in accordance with the Curie-Weiss law upon cooling, then it passes through a broad correlation maximum at about 55 K, characteristic for low-dimensional systems and eventually it is undergoing a kink at $T_N \sim 44$ K, which is replaced by growth. It seems natural to relate the observed anomaly at $T_N$ to establishment of the long-range magnetic order preceding by a short-range one at about 55 K.

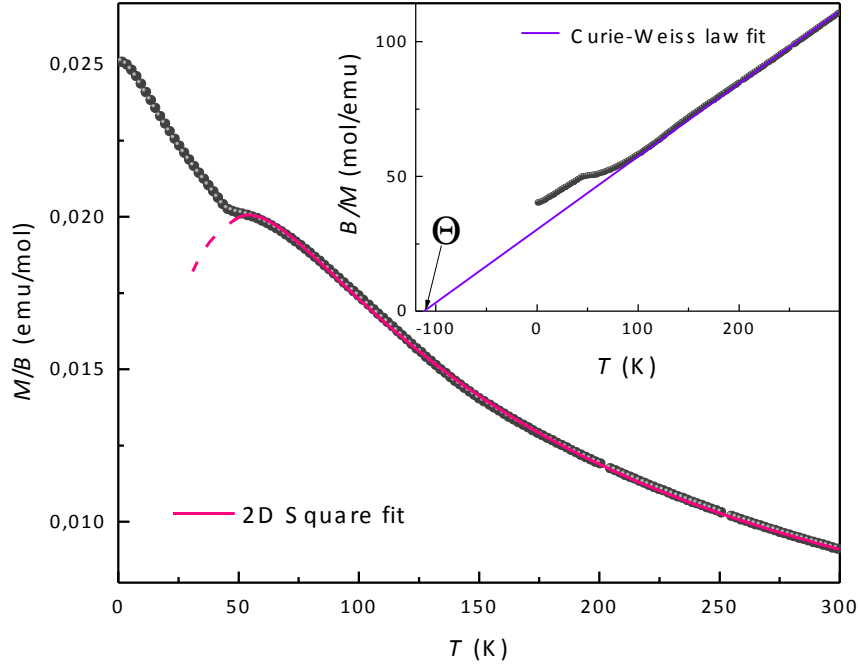

**Fig. 3.** Temperature dependence of the magnetization and its inverse value (in the inset) at $B = 9$ T for NaMnSbO$_4$ recorded in the FC regime. The solid pink curve is an approximation in accordance with 2D square lattice model (eqn.2), the violet line is Curie-Weiss law fit.

The approximation of $\chi(T)$ in accordance with Curie-Weiss law ($\chi = \chi_0 + C/(T - \Theta)$) in the range $250 < T < 300$ K gives $\chi_0 \sim -6.9 \times 10^{-5}$ emu/mol, which is in satisfactory agreement with the value of $\chi_{dia} \sim -7 \times 10^{-5}$ emu/mol obtained by direct summation of Pascal constants [54] for the diamagnetic contributions of atoms constituting NaMnSbO$_4$. The negative Weiss temperature $\Theta \sim -115 \pm 1$ K indicates predominance of the antiferromagnetic interactions in the paramagnetic phase. The effective magnetic moment $\mu_{eff} = \sqrt{3k_B C/\mu_B^2 N_A}$ ($k_B$ is the Boltzmann constant, $C$ is Curie constant and $N_A$ is the Avogadro number) estimated as 5.98 $\mu_B$/f.u. that is in good agreement with expected value for high-spin Mn$^{2+}$ ($\mu_{theor} = \sqrt{g^2 \mu_B^2 S(S+1)} = 5.96 \mu_B$/f.u. using the effective g-factor of 2.01 ± 0.01 determined from ESR data).

Taking into account the structural conditions, which allow to consider the magnetic lattice as almost 2D square, one can account the behaviour of the magnetic susceptibility on the whole temperature range higher than $T_N$ in the framework of the 2D square lattice model. Therefore, it is possible to analyse the dependence $\chi(T)$ using high temperature series expansion (HTE) equation with a temperature-independent term $\chi_0$ [55]:

$$\chi(T) = \frac{Ng^2\beta^2/kJ}{\frac{3T}{JS(S+1)} + \sum_{n=1}^{\infty}\frac{C_n}{[T/JS(S+1)]^{n-1}}} \qquad (1)$$

which can be rewritten by substituting the parameters $C_n$ for the system of spins $S = 5/2$ from work of Lines [55]:

$$\chi(T) = \chi_0 + \frac{0.374g^2/J}{\frac{12T}{35J} + 4 + C_{n1}\frac{J}{T} + C_{n2}\left(\frac{J}{T}\right)^2 + C_{n3}\left(\frac{J}{T}\right)^3 + C_{n4}\left(\frac{J}{T}\right)^4 + C_{n5}\left(\frac{J}{T}\right)^5} \qquad (2)$$

Where $J = J/k$, for the obtained result in Kelvin units and $C_{n1} = 12.67$, $C_{n2} = 17.45625$, $C_{n3} = 175.51953125$, $C_{n4} = 697.55615234375$, $C_{n5} = 871.9451904296875$. The best fit of the $\chi(T)$ experimental data (solid pink line on Fig. 3) according to eqn. 2 resulted in the value $J = -5.3$ K and $g = 1.98$. Both spin exchange parameter $J$ and $g$-factor are in good agreement with density functional analysis ($J_1^{theor} = -5.35$ K at $U_{eff} = 5$ eV) and ESR data ($g^{ESR} = 2.01 \pm 0.01$), respectively (see below). As one can see from Figure 4, the position of anomaly at $T_N \approx 44$ K shifts slightly towards the high-temperature side with increasing magnetic field up to 9 T.

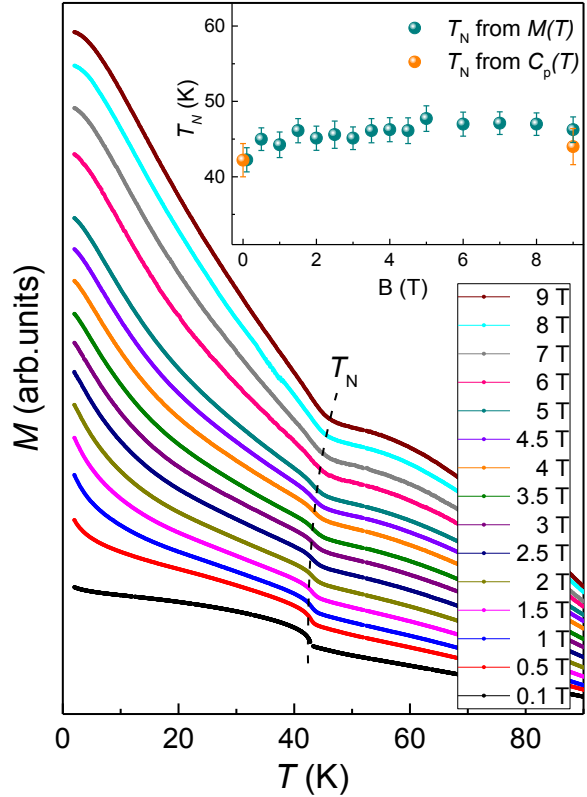

**Fig. 4.** $M(T)$ curves for NaMnSbO$_4$ at various external magnetic fields. The data are arbitrarily shifted along the vertical axes. The dashed curve shows the position of the anomaly $T_N$. Inset: The $T_N$ versus $B$ dependence.

Specific heat data in zero magnetic field are in good agreement with the temperature dependence of magnetic susceptibility, and demonstrate a distinct $\lambda$-shaped anomaly, which directly confirms an onset of the magnetic order at $T_N$ (Fig. 5a). In applied magnetic field the maximum at $T_N$ on the $C_p(T)$ curve broadens and shifts to higher temperatures (inset in Fig. 5a). According to a classical Dulong-Petit law saturation value in NaMnSbO$_4$ is expected to be $C_V = 3Rv = 175$ J/(mol K), where $R$ = 8.31 J/mol K is the gas constant and $v$ = 7 is the number of atoms per formula unit.

In order to analyse the nature of the magnetic phase transition and to evaluate the corresponding contribution to the specific heat and entropy, $C_p(T)$ dependence was also measured for the isostructural non-magnetic compound Na$_2$TeO$_6$ (Fig. 5a). We assume that the specific heat of the isostructural compound Na$_2$TeO$_6$ provides an estimate for the pure lattice counterpart to $C_p(T)$. The correction to this contribution for NaMnSbO$_4$ has been made taking into account the difference between the molar masses for each type of atom in the compound (Na–Mn and Te–Sb) [56]. The values for Debye temperature $\Theta_D$ have been determined as about 343 ± 5 K for the diamagnetic compound Na$_2$TeO$_6$ and 327 ± 5 K for the NaMnSbO$_4$ respectively. It was found that the best description of the specific heat data for non-magnetic Na$_2$TeO$_6$ is achieved by the sum of the Debye ($C_D$) and Einstein ($C_E$) contributions [56]

$$C_{ph} = \alpha C_D(T, \theta_D) + (1-\alpha) \sum_i C_{Ei}(T, \theta_{Ei}) \tag{3}$$

Here $\Theta_D$ and $\Theta_{Ei}$ are the Debye and Einstein temperature, respectively, and $\alpha$ denotes a relative weight factor for Debye and Einstein terms. The fit according to eqn. 3 resulted in the values $\Theta_{E1}$ = 120± 5 K, $\Theta_{E2}$ = 145± 5 K and $\Theta_{E3}$ = 240± 5 K (Fig. 5a).

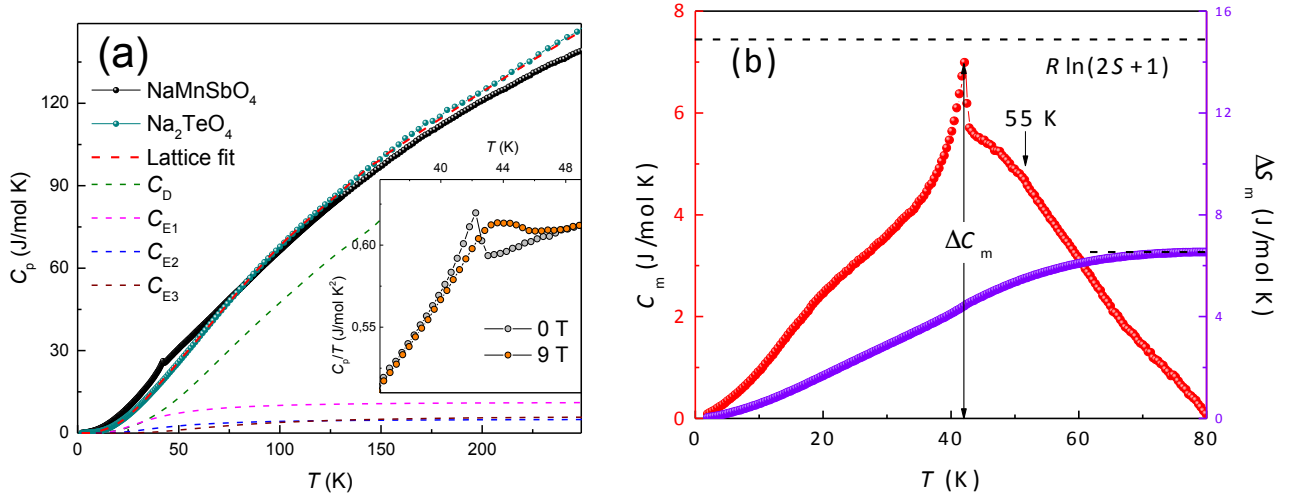

**Fig. 5.** (a) Temperature dependencies of the specific heat of NaMnSbO$_4$ (black symbols) and its non-magnetic isostructural analogue Na$_2$TeO$_6$ (green symbols) in zero magnetic field; the dashed line presents the determined lattice contribution. (b) Magnetic specific heat $C_m$ (red balls) and magnetic entropy $S_m$ (violet balls) variation. Inset: Zoomed-in low temperature part of $C_p$ at $B$ = 0 (grey symbols) and $B$ = 9 T (orange symbols).

A jump of the specific heat at the $T_N$ $\Delta C_m \sim 7$ J/(mol K) is lower than the estimation from the mean-field theory $\Delta C_m = 5R \cdot S(S + 1)/[S^2 + (S + 1)^2] \approx 19.7$ J/(mol K) [56] (Fig. 5b). It is important to note that the correlation maximum observed on the $\chi(T)$ dependencies at about $T_{max} \sim 55$ K is also quite obvious on the $C_m(T)$, but its position slightly shifts to lower temperature. The corresponding value of magnetic entropy saturates at about 60 K reaching approximately 6.5 J/(mol K). It is also lower than the estimate obtained from the mean-field theory $\Delta S_m = R\ln(2S + 1) \approx 14.9$ J/(mol K). One should note that the magnetic entropy released below $T_N$ removes only about 30% of the theoretical saturation value. Such a reduction of the $\Delta C_m$ and the $\Delta S_m$ indicates the presence of strong short-range magnetic correlations at temperatures higher the order-disorder transition, which is typical of low-dimensional compounds [4, 9].

The full magnetization isotherms for NaMnSbO$_4$ at various temperatures are presented in Fig. 6. The magnetic hysteresis loop in combination with the linear M-B behaviour at high-field is a feature of a canted AFM ordering [57-59]. The hysteresis loop narrows with increasing temperature and eventually disappears at $T > T_N$ (Fig. S1 [46]), confirming that it arises from a long-range order. This feature is intrinsic property of the system under study. In contrast, the $M(B)$ curves in the vicinity the short-range ordering regime ($T \approx T_{max}$) pass exactly through the original point. Nevertheless, despite the absence of a hysteresis loop $M(B)$ still demonstrates a small spontaneous moment that decays slowly. Eventually, the linear behaviour is observed at $T > 100$ K, typical of paramagnets or robust antiferromagnets with strong exchange interactions. To estimate the spin canting angle, we determine by extrapolation the y-axis intercept of $M(0) = 0.081\mu_B$, the canted moment at zero field (Fig. 6). The corresponding canting angle $\varphi$ can be calculated using the expression $\varphi = \sin^{-1}(M(0)/gS\mu_B)$ [58]. Using $S = 5/2$ and $g = 2$ for Mn$^{2+}$ ions, we obtain $\varphi \approx 1°$.

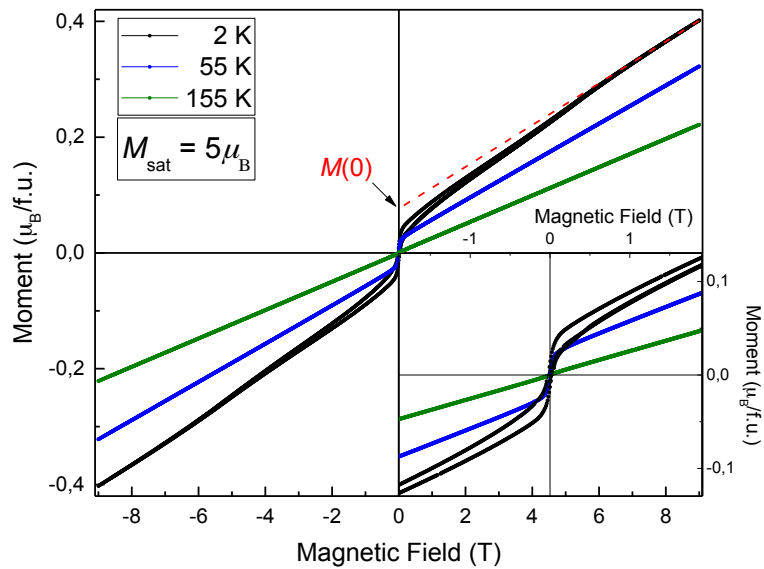

**Fig. 6.** Field dependences of magnetization $M(B)$ of NaMnSbO$_4$ at various temperatures. The red dashed line denotes the linear fit of the $M(B)$ for $B > 7$ T at 2 K. Inset: the zoomed-in central parts of $M(B)$.

No signature of any other field-induced anomalies was evidenced up to 9 T. One can note that within the investigated range of the applied magnetic fields, the magnetic moment is still far below the theoretically expected saturation magnetic moment $M_S = gS\mu_B = 5\mu_B$ for $Mn^{2+}$ ($S = 5/2$) ions.

It is very interesting that the shape of hysteresis loop is quite unusual. It is squeezed in the middle, often denoted as a "constricted loop" or "wasp-waisted hysteresis loop". No convincing account is generally given on this phenomenon at the moment, because it can be associated with several different physical causes. Possible reported explanations were listed for example in Ref. 60. Among them there are the influence of dipolar interactions, the effect of surface anisotropy superimposed to bulk anisotropy, bimodal coercivity distribution, the mixing of competing anisotropies (uniaxial and cubic), etc.

Such type of the hysteresis was observed for many different magnetic materials: for example, for $Dy_{2-x}Eu_xTi_2O_7$ pyrochlore [61], for monodispersed ferromagnetic FePt nanoparticles [60], for soft-magnetic NiCoP-coated hard-magnetic M-type ferrite $BaFe_{12}O_{19}$ polystyrene (PS) bilayer composite films [62], for $Fe_2O_3$-$SiO_2$ nanocomposite [63], manganite/ruthenate $La_{0.67}Sr_{0.33}MnO_3$/$SrRuO_3$ superlattices [64], for one-dimensional (1D) nanotubes of $Nd_{0.1}Bi_{0.9}FeO_3$ (NBFO) [65], for a single magnetic phase $CoFe_2O_4$ nanopowder [66].

Quite close for us example with similar type of hysteresis loop has also been observed for 2D multiferroic compound $BaNiF_4$ [67]. According to the theoretical calculations this system is canted antiferromagnet with weak interaction between different sheets [68], leading to low magnetic ordering temperature and pronounced two-dimensional behaviour. $BaNiF_4$ was related to 2D magnetic vortex systems, for which the wasp-waisted type of $M$-$B$ curves was declared [69]. In this case the spins tend to align circle wise at low magnetic fields and parallel to the magnetic field when the field is large enough. Ederer and Spaldin pointed that the AFM aligned net spins due to spin canting are easy to be aligned at high enough field, leading to the drastically increased magnetization and weak ferromagnetism. However, at low field, due to the weak AFM nature of canted spin, the net spins due to canting will turn back to their initial orientation and weak antiferromagnetism is recovered, leading to the drastic decrease of magnetization and the wasp-waisted type $M$-$B$ curve.

## C. Dynamic magnetic properties

Over the whole temperature range investigated the ESR spectra reveal the presence of exchange-narrowed Lorentzian shaped resonance mode with g-factor close to 2, which is naturally to assign to the signal from S-type $Mn^{2+}$ ion in octahedral oxygen coordination. At temperatures below $T < 50$ K the degradation of the ESR signal occurs, indicating onset of long-range order and corresponding opening of the energy gap for resonance excitations. The representative ESR spectrum of powder sample $NaMnSbO_4$ at room temperature is presented on the inset in Fig. 7. An accurate analysis of the lineshape requires including two circular components of the exciting linearly polarized microwave field into the fit formula since the line observed is relatively broad (only one order less than the resonance field in the present compound) [70]:

$$\frac{dP}{dB} \propto \frac{d}{dB}\left[\frac{\Delta B}{\Delta B^2 + (B - B_r)^2} + \frac{\Delta B}{\Delta B^2 + (B + B_r)^2}\right] \quad (4)$$

This is a symmetric Lorentzian line, where $P$ is the power absorbed in the ESR experiment, $B$ is magnetic field, $B_r$ is resonance field, $\Delta B$ is the linewidth. It was found that in the whole $T$-range experimental data can be fitted by single Lorentzian in a form of (4). Result of ESR lineshape fitting is shown by blue solid line in Fig. 7. Evidently, the fitted curves in reasonable agreement with the experimental data.

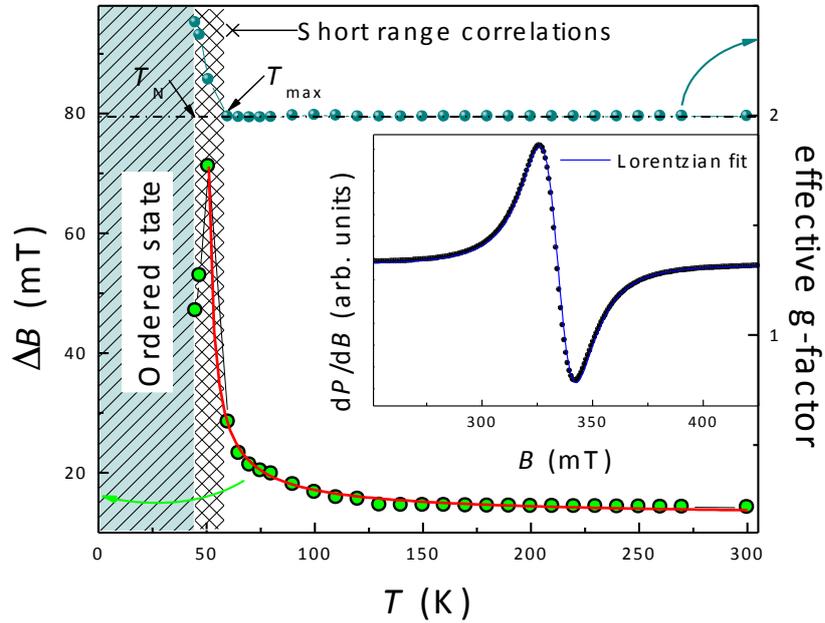

**Fig. 7.** The temperature dependencies of the the ESR linewidth $\Delta B$ and effective g-factor of NaMnSbO$_4$. Solid red curve is the approximation according eqn. 5. On the inset: the representative ESR spectrum taken at room temperature: black symbols are experimental data, solid blue line is fitting by Lorentzian (eqn. 4)

The parameters of the ESR spectra obtained from the approximation are also collected in Fig. 7. The effective $g$-factor remains almost temperature-independent down to 50 K, then the rapid shift of the resonance field occurs on approaching the critical temperature $T_N \sim 44$ K from above. Such behaviour indicates the establishment of a long-range magnetic order and appearance of internal magnetic field which violates the resonance conditions.

Broadening of the ESR line with decreasing temperature has been observed earlier for the wide class of antiferromagnetic, spin-glass and diluted magnetic systems [71-76] and explained in the frame of theory of critical behaviour of ESR linewidth in the vicinity of the order-disorder transition. According to Mori-Kawasaki-Huber theory, the temperature dependence of the ESR linewidth $\Delta B$ can be described by [77]:

$$\Delta B(T) = \Delta B^* + A \cdot \left[ \frac{T_N^{ESR}}{T - T_N^{ESR}} \right]^\beta \qquad (5)$$

where the first term $\Delta B^*$ describes the high-temperature exchange narrowed linewidth, which is temperature independent, while the second reflects the critical behaviour with $T_N^{ESR}$ being the critical temperature and $\beta$ the critical exponent.

An approximation of the ESR spectra has shown that $\Delta B(T)$ can be reasonably well described in the frame of this model (Eqn. 5) over a wide temperature range 50 - 300 K (red solid curve on Fig. 7). The best fitting was attained with the parameters $\Delta B^* = 12 \pm 1$ mT, $T_N^{ESR} = 48 \pm 1$ K and $\beta = 0.80 \pm 0.10$. Obviously the $T_N^{ESR}$ value is close to the critical temperature $T_N$, estimated from the thermodynamic data.

The value of the critical exponent $\beta = 0.8$ is noticeably larger than the expected $\beta = 1/3$ for the 3D Heisenberg antiferromagnet in the frame of Kawasaki approach [77] and probably indicates rather 2D character of exchange interactions in NaMnSbO$_4$ in accordance with quasi-2D network of magnetic sublattice. For example, such values were observed earlier for other 2D antimonates of alkali and transition metals, such as $\beta \sim 0.9$ for 2D Li(Na)$_3$Ni$_2$SbO$_6$ [78], $\beta \sim 0.6$ for Li$_4$FeSbO$_6$ [6].

The nuclear magnetic resonance study was performed using $^{23}$Na nuclear which has spin 3/2 and quadrupole moment $10.4 \times 10^{-30}$ m$^2$. Left panel of Fig. 8 presents a spectrum obtained at 150 K, which can be reasonably described by the Hamiltonian

$$\hat{H} = -\gamma_N \hbar I_Z B_{ext} + \hat{I}^i A_{ij} \hat{S}^j + \frac{e^2 qQ}{4I(2I-1)} \left[ 3\hat{I}_Z^2 - \hat{I}^2 + \frac{\eta}{2}\left(I_+^2 + I_-^2\right) \right] \qquad (6)$$

where the first term represents the Zeeman interaction with the external magnetic field, the second term is an anisotropic transferred hyperfine coupling of the nuclear spin with the electron spins, and the third term is the quadrupolar coupling (here $\gamma_n$ is the gyromagnetic ratio of the nucleus, $I$ the nuclear spin, $S$ the electron spin, $B_{ext}$ the external magnetic field, $A_{ij}$ the hyperfine coupling tensor, $\hbar$ the reduced Planck constant, $e$ the elementary charge, $Q$ the nuclear electric quadrupole moment. $q$ and $\eta$ are respectively the electric field gradient (EFG) main value and the asymmetry parameters, defined as $eq=V_{zz}$ and $\eta=V_{yy}-V_{xx}/V_{zz}$.) [79]. The simulation of the spectrum of a powder sample at a relatively high temperature 150 K using Hamiltonian (6) was done according to the procedure described, for example, in [80]. The quadrupolar frequency $\omega_Q = 3e^2 qQ/[2I(2I-1)\hbar]$ and the EFG asymmetry parameter $\eta$ were taken as fitting parameters. The best result was obtained with the values $\omega_Q = 2$ MHz and $\eta = 0.36$. The symmetry of hyperfine tensor (also taken as a fitting parameter) was found to be almost axial. The spectrum gradually shifts and broadens with decreasing temperature, then begins to transform to rectangular-like shape in the vicinity of the magnetic order transition temperature (right panel of Fig. 8). Dynamic properties of the electron spin system were studied by measurements of the longitudinal nuclear relaxation rates $T_1^{-1}$ at the maximum of the spectral intensity in the external field magnetic field of 3.67 T within temperature range from 300 K down to 35 K. A typical delay time dependence of the longitudinal nuclear magnetization $M(t)$ for $T = 77$ K is shown in the inset of Figure 9. For a spin-3/2 nucleus this recovery measured in main transition +1/2 − -1/2 is expected to follow a two exponential law [81]:

$$M(\tau) = M_0 \left(1 - p\left(0.9\exp\left(-\left(\frac{6\tau}{T_1}\right)^b\right) + 0.1\exp\left(-\left(\frac{\tau}{T_1}\right)^b\right)\right)\right) \quad (7)$$

Where $\tau$ is a delay after series of saturating pulses, $T_1$ is the relaxation time of $^{23}$Na nuclei, $p$ is a saturation parameter which is close to 1. Parameter $b$ is the stretching exponent parameter, which reflects a distribution of fluctuation frequencies and spatial anisotropy of the localized electron spin subsystem [82-87] (in our experiments $b$ decreases with temperature within the range $0.5 < b < 0.9$). There is no pronounced peak but a small kink of the relaxation at $T_N$ which is often found in low-dimensional compounds. The temperature dependence of the $^{23}$Na spin-lattice relaxation rate corresponds to the Moriya formula [88]:

$$T_1^{-1} = T_{1\infty}^{-1}\left[1 + \frac{J^*}{4k_BT} + n\frac{1}{2}\left(\frac{J^*}{k_BT}\right)^2\right]\exp\left[-\frac{1}{2}\left(\frac{J^*}{k_BT}\right)^2\varsigma\left(1 + \frac{J^*}{4k_BT}\right)\right] \quad (8)$$

where $J^* = -2J$. The term involving $(J^*/k_BT)^2$ in the prefactor has a negligible effect at $T >> J^*$ $\varsigma = (z/6)\cdot[S(S + 1)] \approx 6$ for the electron spin $S = 5/2$ and the number of nearest neighbors of each spin $z = 4$ for 2D square lattice. From the fitting we obtain $J = -5.927 \pm 0.556$ K, which is in a good agreement with the results of magnetization and DFT studies (sec. III.B and III.D).

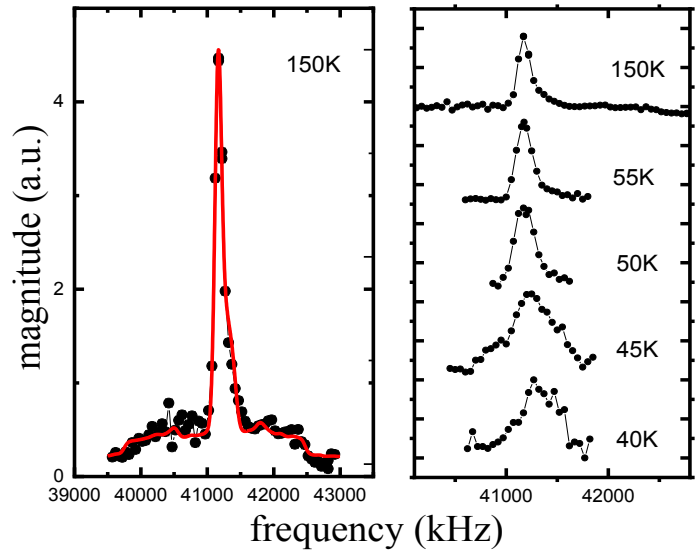

**Fig. 8.** (left)$^{23}$Na NMR spectra: black circles are experimental data, solid line is fitting by Hamiltonian (eqn. 6) with powder averaging. (right) Temperature evolution of the spectra in temperature range 40 – 150 K.

In general, nuclear magnetic spin-lattice relaxation is dominated by fluctuating fields at the nuclear site produced by electron spins:

$$\frac{1}{T_1} \sim \gamma_n^2 k_B T \sum_q A(\hat{q}) \frac{\chi''(\hat{q},\omega_L)}{\hbar \omega_L} \qquad (9)$$

where the sum is over the wave vectors $q$ within the first Brillouin zone, $\chi''(q,\omega_L)$ is the imaginary part of the dynamic susceptibility, $A(q)$ the $q$-depended nuclei form factor proportional to the perpendicular component of hyperfine tensor, $\omega_L$ the Larmor frequency, $\gamma_n$ the nuclear gyromagnetic ratio and $k_B$ the Boltzmann constant. Therefore, the term $1/(T_1 T)$ is proportional to the local dynamic susceptibility of a magnetic subsystem measured at the nucleus position at the Larmor frequency. In the paramagnetic regime, where the electronic correlations are weak, and the fluctuation rate is significantly far from the Larmor frequency, the dependence of the susceptibility on the $q$-vector and frequency can be neglected. Thus, $1/(T_1 T)$ probed by the $^{23}$Na NMR at 3.67 T is expected to be proportional to the static bulk susceptibility obtained at 3.5 T, and this linear dependence is observed at temperature range 300 - 77 K. Below 70 K linear dependence disappears, indicating the developing the correlations within a short-range ordering region and the critical slowing down of the magnetic fluctuations in the vicinity of Néel temperature (Fig. 10). The temperature region where the correlations develops, observed by NMR, is in good agreement with the region where the bulk susceptibility deviates from the Curie-Weiss law.

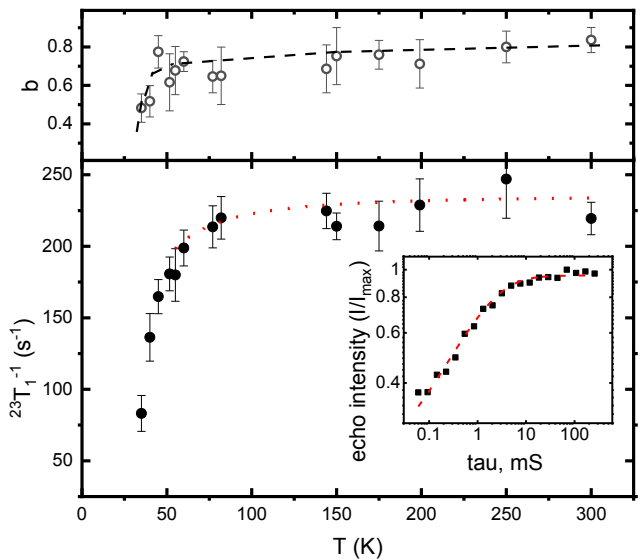

**Fig. 9.** Main panel: the temperature dependence of $^{23}$Na nuclear spin-lattice relaxation rate: black circles are experimental data, dotted line is fitting as described in Ref. 88. Upper panel: temperature dependence of the stretched exponent parameter $b$ (black dashed line is guide for the eye); Insert: recovery of spin-echo intensity plotted as a function of delay between pulses. Dashed line represents the fit to eqn. 7.

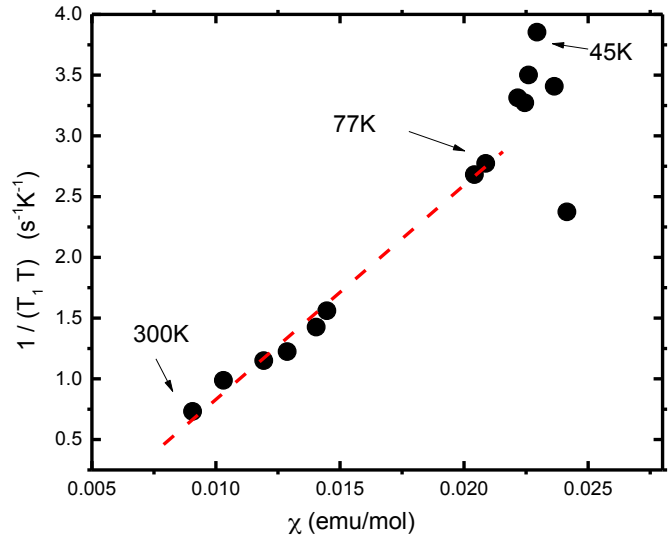

**Fig. 10.** Local dynamical susceptibility $1/(T_1T)$ (left y-axis, black circles) plotted as a function of the static bulk susceptibility measured at 3.5 T. The dashed red line marks the region, where dynamic susceptibility follows to the static one.

## D. Density functional analysis of magnetic structure

To gain insight into the observed magnetic properties of NaMnSbO$_4$, we have determined its spin exchange interactions by carrying out energy-mapping analysis based on DFT calculations [89-91]. The projection view of the crystal structure of NaMnSbO$_4$ along the *a* axis is given in Figure 11a, which shows that the layers of Mn$^{2+}$ (d$^5$, $S = 5/2$) ions alternate with those of the Sb$^{5+}$ ions along the *b* axis. A projection view of a layer of Mn$^{2+}$ (d$^5$, $S = 5/2$) ions along the b axis is presented in Figure 11b.

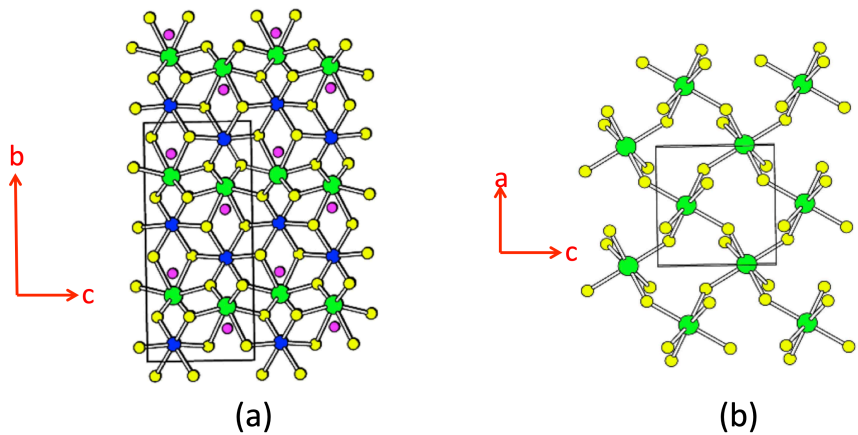

**Fig. 11.** (a) Projection view of NaMnSbO$_4$ along the *a* axis. (b) Projection view of an isolated layer made up of corner-sharing MnO$_6$ octahedra. Green circles: Mn, Blue circles: Sb, yellow circles: O, and Magenta circles: Na.

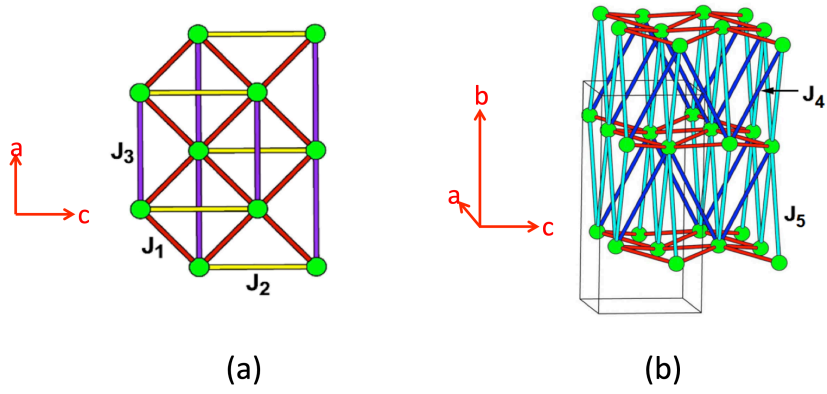

(a)                      (b)

**Fig. 12**. (a) Spin exchange paths $J_1 - J_3$ within a layer of $Mn^{2+}$ ions parallel to the *ac*-plane. (b) Spin exchange paths $J_4$ and $J_5$ between adjacent layers of $Mn^{2+}$ ions.

We consider three spin exchange paths $J_1$, $J_2$ and $J_3$ within each layer of $Mn^{2+}$ ions (Fig. 12a), and two spin exchange paths $J_4$ and $J_5$ between adjacent layers of $Mn^{2+}$ ions (Fig. 12b).

To determine the values of $J_1 - J_5$, we carry out spin-polarized DFT calculations using the (2*a*, 2*b*, *c*) supercell for $NaMnSbO_4$ containing 16 formula units (FUs) for the six ordered spin states (FM, AF1, AF2, AF3, AF4, AF5) defined in Figure 13. The calculations were performed by employing the projected augmented wave method encoded in the Vienna Ab Initio Simulation Package (VASP) [92-94] with the generalized gradient approximation of Perdew, Burke and Ernzerhof [95] for the exchange-correlation functionals with a plane wave cutoff energy of 450 eV, a set of 4×4×6 k-points, and a threshold $10^{-6}$ eV for energy convergence. The DFT plus on-site repulsion $U$ (DFT+U) method [96] was employed at $U^{eff} = U - J = 3$, 4 and 5 eV to describe the electron correlation in the Mn 3d states. Given the spin Hamiltonian,

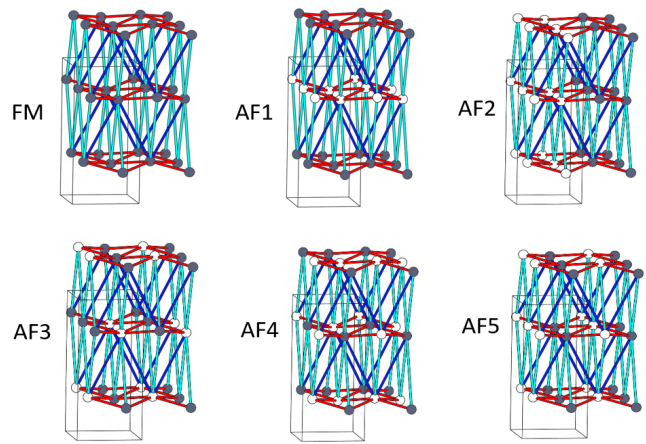

**Fig. 13.** Six ordered spin arrangements of $NaMnSbO_4$ employed to extract the values of the spin exchanges $J_1 - J_5$ by energy-mapping analysis, where the grey and white circles indicate the up and down spins of $Mn^{2+}$ ions. For the simplicity, only three spin exchange paths, $J_1$, $J_4$ and $J_5$ are indicated in red, blue and cyan cylinders, respectively.

$$H_{spin} = -\sum_{i<j} J_{ij}\vec{S}_i \cdot \vec{S}_j, \tag{10}$$

where $\vec{S}_i$ and $\vec{S}_j$ are the spins at magnetic ion sites $i$ and $j$, respectively, with the spin exchange constant $J_{ij} = J_1 - J_5$.

Then, for each ordered spin state, the total spin exchange energy per 16 FUs can be expressed as

$$E = (n_1 J_1 + n_2 J_2 + n_3 J_3 + n_4 J_4 + n_5 J_5)(N^2/4), \tag{11}$$

where $N$ is the number of unpaired spins per $Mn^{2+}$ (i.e., $N = 5$), and the coefficients $n_1 - n_5$ for the six ordered states are summarized in Table S1 of the Supplemental Materials [46]. The relative energies of the six ordered spin states determined by DFT+U calculations are summarized in Table 4. Then, by mapping these relative energies onto the corresponding ones in terms of the total spin exchange energies, we obtain the values of $J_1 - J_5$ listed in Table 5. The latter reveals that all spin exchanges are AFM, and that the side exchange $J_1$ dominates over all other spin exchanges. Table 5 also lists the Curie-Weiss temperatures $\Theta_{cal}$ calculated on the basis of the calculated spin exchanges $J_1 - J_5$ using the mean-field approximation [97]. The $\Theta_{cal}$ changes from -113 to -92 to -68 K as $U^{eff}$ increases from 3 to 4 to 5 eV, respectively. The $\Theta_{cal} \approx -113$ K obtained with $U^{eff} = 3$ eV are in good agreement with the experimental value of -112 ± 1 K. Therefore, our calculations predict that each layer of $Mn^{2+}$ ions would be antiferromagnetically ordered without much spin frustration, and then these AFM layers will be antiferromagnetically ordered leading to a three-dimensional (3D) AFM state. At the same time non-trivial character of the anomaly on the temperature-dependent magnetic susceptibility and wasp-waisted hysteresis loop may indicate the more complicated probably canted antiferromagnetic state. In order to clear up the ground state the neutron studies are very desirable.

**Table 4**. Relative energies (in meV per 16 FUs) of the spin ordered states determined by DFT+U calculations.

| State | $U_{eff} = 3$ eV | $U_{eff} = 4$ eV | $U_{eff} = 5$ eV |
|---|---|---|---|
| FM | 21.4 | 15.9 | 11.8 |
| AF1 | 20.6 | 15.3 | 11.3 |
| AF2 | 10.4 | 7.7 | 5.7 |
| AF3 | 0 | 0 | 0 |
| AF4 | 9.7 | 7.2 | 5.3 |
| AF5 | 0 | 0 | 0 |

**Table 5**. Values (in $k_B$K) of the spin exchanges $J_1 - J_5$ and the Curie-Weiss temperature (in K) determined by DFT+U calculations.

|  | $U_{eff}$ = 3 eV | $U_{eff}$ = 4 eV | $U_{eff}$ = 5 eV |
|---|---|---|---|
| $J_1$ | -9.72 | -7.24 | -5.35 |
| $J_2$ | -0.29 | -0.22 | -0.18 |
| $J_3$ | -0.56 | -0.42 | -0.33 |
| $J_4$ | -0.37 | -0.29 | -0.23 |
| $J_5$ | -0.19 | -0.14 | -0.11 |
| $\Theta_{cal}$(K) | -112.8 | -91.7 | -68.1 |

## IV. CONCLUSION

NaMnSbO$_4$ is a new orthorhombic phase based on a distorted double-layered hexagonal close packing of oxide anions with cations filling 3/4 octahedral voids in an ordered manner. Its magnetic sublattice is a pseudo-tetragonal bidimensional array of MnO$_6$ octahedra sharing vertices, with Mn-O-Mn angles of 128.0°.

When the temperature decreases the magnetic susceptibility demonstrates the behaviour typical for low-dimensional magnets with broad correlation maximum at $T_{max}$ = 55 K followed by clear anomaly at $T_N \approx$ 44 K, which can be assigned to the long-range magnetic ordering. At the same time unusual character of the $T_N$ anomaly and the wasp-waisted hysteresis loop on the magnetization curve indicate that the long-range state is most probably canted antiferromagnetic.

ESR spectra of NaMnSbO$_4$ in paramagnetic phase reveal Lorentzian shape signal, which can be attributed to Mn$^{2+}$ ions and characterized by the temperature independent effective g-factor $g$ = 2.01 ± 0.1. ESR data indicate an extended region of short-range order correlations, typical of low-dimensional or frustrated magnets. An analysis of the ESR line critical broadening performed in the frame of the Kawasaki-Mori-Huber theory confirms low-dimensional character of magnetic exchange interactions in the compound under study.

DFT calculations have shown that ground state is stabilized by the five exchange parameters. It was found that all spin exchange are AFM, and that the side exchange $J_1$ on square dominates over all other spin exchanges indicating rather 2D magnetism in new NaMnSbO$_4$ compound.

Temperature dependence of the magnetic susceptibility and $^{23}$Na nuclear spin lattice relaxation are described reasonably well in the framework of 2D square lattice model with the main exchange parameter $J$ = -5.3 K, which is in good agreement with density functional analysis and ESR data.

**Supplemental Materials:** crystallographic information file (cif), details of sample preparation, EDX analysis, and values of the coefficients $n_1 - n_5$ in Eq. 11 (pdf).

ACKNOWLEDGEMENTS

The X-ray study by V.B.N and I.L.S. was supported by the Russian Foundation for Basic Research (grant 11-03-01101) and the International Centre for Diffraction Data (grant-in-aid 00-15); magnetic studies by T.M.V., E.A.Z. and A.N.V. were supported by Russian Foundation for Basic Research through grants 17-52-45014 and 18-02-00326 and by the Russian Ministry of Education and Science of the Russian Federation through NUST «MISiS» grant K2-2017-084 and by the Act 211 of the Government of Russia, contracts 02.A03.21.0004 and 02.A03.21.0011. The work at KHU was supported by Basic Sciences Research Program through the National Research Foundation of Korea (NRF) funded by the Ministry of Education (NRF-2017R1D1A1B03029624). ). The authors are thankful to Dr.Yu.V. Popov for the EDX measurements.